# AV-Net: Deep learning for fully automated artery-vein classification in optical coherence tomography angiography

**MINHAJ ALAM,**[1,3] **DAVID LE,**[1,3*] **TAEYOON SON,**[1] **JENNIFER I. LIM,**[2] **AND XINCHENG YAO**[1,2,*]

[1]*Department of Bioengineering, University of Illinois at Chicago, Chicago, IL 60607, USA*
[2]*Department of Ophthalmology and Visual Sciences, University of Illinois at Chicago, Chicago, IL 60612, USA*
[3]*These authors contributed equally to this work*
*\*xcy@uic.edu*

**Abstract:** This study is to demonstrate deep learning for automated artery-vein (AV) classification in optical coherence tomography angiography (OCTA). The AV-Net, a fully convolutional network (FCN) based on modified U-shaped CNN architecture, incorporates enface OCT and OCTA to differentiate arteries and veins. For the multi-modal training process, the enface OCT works as a near infrared fundus image to provide vessel intensity profiles, and the OCTA contains blood flow strength and vessel geometry features. A transfer learning process is also integrated to compensate for the limitation of available dataset size of OCTA, which is a relatively new imaging modality. By providing an average accuracy of 86.75%, the AV-Net promises a fully automated platform to foster clinical deployment of differential AV analysis in OCTA.

## 1. Introduction

Early disease diagnosis and effective treatment assessment are essential to prevent vision loss. Differential artery-vein (AV) analysis can provide valuable information for disease detection and classification. It has been demonstrated to be valuable for evaluating diabetes, hypertension, stroke and cardiovascular diseases [1-3] along with common retinopathies [4, 5]. Several clinical studies have evaluated AV abnormalities in different diseases. However, clinical deployment of the AV analysis for routine management of eye diseases is challenging. Most of the clinical studies relied on manual or semi-automated approaches to identify arteries and veins, which is ineffective in a clinical setting. Therefore, a fully automated platform for AV classification is important.

To date, automated AV classification has been primarily used in color fundus images acquired with traditional fundus photography [6-15], which provide limited resolution and sensitivity to reveal microvascular abnormalities associated with eye conditions [16]. Microvascular anomalies that occur at early stages of eye diseases, cannot be reliably identified in traditional fundus photography [17-19]. An alternative to traditional color fundus imaging is optical coherence tomography (OCT) and OCT angiography (OCTA). OCT and OCTA can provide depth-resolved visualization of individual retinal layers with capillary level resolution. Especially, OCTA is sensitive to identify subtle microvascular changes, and thus have been extensively explored for quantitative analysis and objective classification of retinal diseases [20-24]. Using quantitative feature analysis, we have recently demonstrated the potential of differentiating artery and vein in OCTA [4, 5, 25, 26]. Differential AV analysis showed improved OCTA performance to identify abnormal changes in diabetic retinopathy (DR) and sickle cell retinopathy (SCR) eyes [4, 5, 26]. However, clinical deployment of the AV analysis in OCTA requires an automated, simple, but robust method. A potential solution could be the employment of deep machine learning i.e., convolutional neural networks (CNNs) for AV classification automatically. A fully convolutional network (FCN) can be trained with a ground truth dataset for a specific task and can be implemented on validation or testing dataset. A fully automated method is a key factor for clinical deployment of artificial intelligence (AI) based screening, diagnosis, and treatment evaluation.

In this study, we present and validate AV-Net, an FCN based on a modified U-shaped CNN architecture, for deep learning based AV classification in OCTA.. A multi-modal training process involves both enface OCT and OCTA, which provide intensity and geometric profiles, respectively, for AV classification. Transfer learning is employed to compensate for the limitation of available dataset size of OCTA which is a relatively new imaging modality. By incorporating transfer learning and multi-modal training approaches, fully automated AV classification is demonstrated. The AV-Net performance is validated with manual AV ground truth maps using accuracy and intersection over union (IOU) metrics.

## 2. Methods

This study is in adherence to the ethical standards present in the Declaration of Helsinki and was approved by the institutional review board of the University of Illinois at Chicago (UIC).

### 2.1 Data acquisition

The dataset comprised of 6 x 6 mm$^2$ field of view (FOV) OCT/OCTA data acquired using ANGIOVUE spectral domain (SD) OCTA systems (Optovue, Fremont, CA) with a 70-kHz A-scan rate, a lateral resolution of ~15 µm, and an axial resolution of ~5µm. The inclusion criteria for quality of OCTA acquisition is quality 6 or greater. All OCTA images were qualitatively examined for severe motion or shadow artifacts. Images with significant artifacts were excluded for this study. OCTA data was exported using ReVue (Optovue).

### 2.2 Model Implementation

In this paper, we present for the first time 'AV-Net', an FCN based on a modified U-Net architecture. The input of the AV-Net is a 2-channel system to combine grayscale enface OCT and OCTA. Enface OCT is a near infrared (NIR) image, which is equivalent to a fundus image, to provide vessel intensity profiles. On the other hand, OCTA contains the information of blood flow strength and vessel geometry features. The output of AV-Net is an RGB (red-green-blue) image, with R and B channels corresponding to arteries and veins, respectively, and the G channel presents the background.

The overall design of the AV-Net follows an encoder-decoder architecture (Fig.1a). The encoder, also known as the contracting path, extracts the context of the image. The decoder, also termed as the expanding path, identify image features. The addition of bridging between the encoder and decoder is to enable precise localization and mapping of feature maps to produce the output image [27]. In AV-Net, the encoder is composed of convolution blocks, dense blocks, and transition blocks (Fig. 1b). The convolution blocks are similar to the identity block in ResNet, except for the use of concatenation instead of summation operations [28, 29]. The dense block is composed of convolution blocks, with each subsequent block connected to the previous blocks by skip-connections.

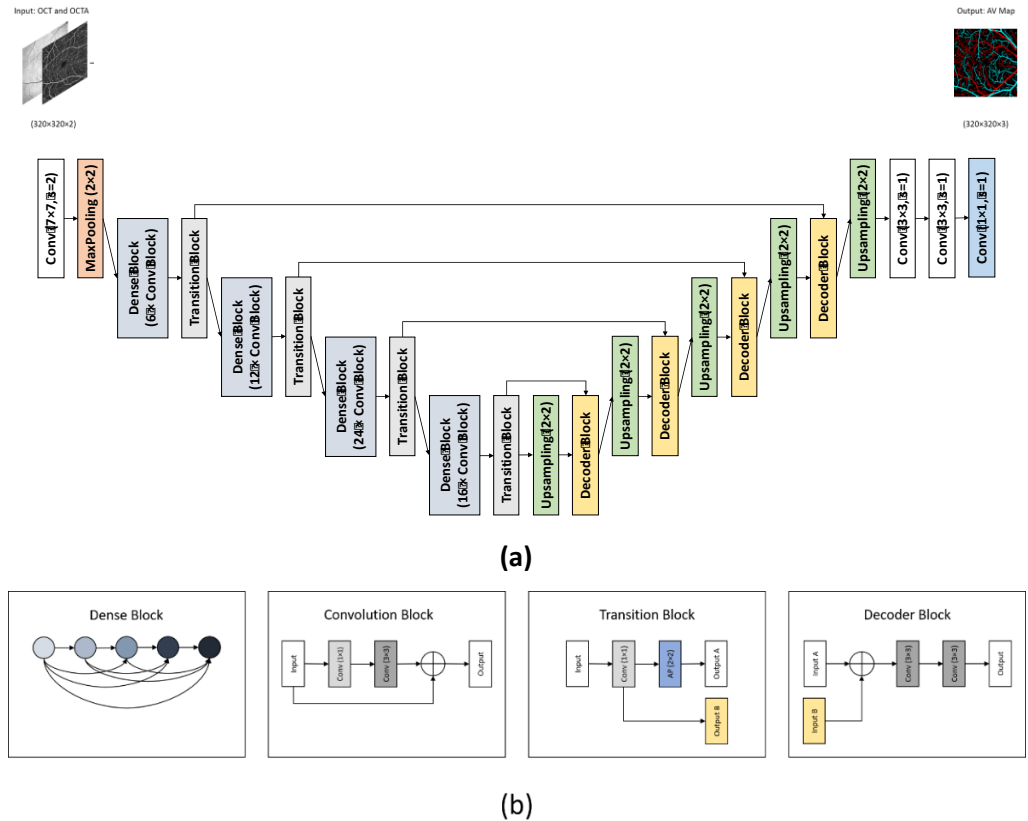

Figure 1. Network architecture for AV-Net, (a) overview of the blocks in AV-Net architecture, (b) the individual blocks that comprises AV-Net. In this figure, Conv stands for convolution operations, AP stands for Average Pooling operation. Each transition block has two outputs, Output A is the output of the AP operation, and Output B is the output of the Conv operation. The skip-connections from each transition block are Output B. In the decoder block, the Input A is the output of the preceding layer, whereas Output B is the output of the appropriately sized transition block.

Skip-connections is to alleviate the vanishing-gradient problem in deep learning [30]. Following each dense block, a transition block is used to reduce the dimensions of the output feature maps. In the decoder, we employ upsampling operations and use decoder blocks. The decoder block concatenates the outputs of the upsampling operation and the output of the convolution from the appropriate transition block. The feature maps are then convolved to enable precise localization of image features.

In the AV-Net, all convolution operations are followed by batch normalization and ReLU activation function, whereas the final convolutional layer is followed by a softmax activation function. Transfer learning was implemented for the encoder network from pre-trained weights optimized from the ImageNet Dataset for AV classification in OCTA. FCN training procedure utilized the Adam optimizer with a learning rate of 0.0001, a dice loss function, and a minibatch size of 8. Moreover, regularization procedures including data augmentation and cross-validation were used to prevent overfitting. Training was performed on a Windows 10 computer using NVIDIA Quadro RTX 5000 Graphics Processing Unit (GPU). The FCN was trained and evaluated on Python (v3.7.1) using Keras (2.2.4) with Tensorflow (v1.31.1) backend. In our study, our OCTA dataset comprised of 40 images and to evaluate our network, a 5-fold cross validation method, with each fold following an 80/20 train/test split procedure, was employed. Due to a limited dataset, data augmentation, i.e., random flips, rotation, zooming, and image shifting, was implemented during training. Therefore, in each fold the network was trained with 3,000 images, and testing evaluation was performed on the 8 original images of each fold. Average accuracy, intersection-over-union (IOU) and F1-score was used as an evaluation metric for AV classification, by comparing with manually labelled ground truths.

*2.3 Loss Functions*

In this study, the AV-Net was trained using a compound loss function derived from dice loss [33] and focal loss [34] and was defined as Eq. 1:

$$L = L_{dice} + L_{focal}$$

Where $L_{dice}$ is the dice loss (Eq. 2) and $L_{focal}$ is the focal loss (Eq. 3). Recent studies have found the combination of multiple losses improves image segmentation tasks with class imbalances [31, 32]. Dice score measures the degree of overlap between the prediction and ground truth and is therefore suited for image segmentation (pixel-wise classification) tasks. The dice loss can be written as

$$L_{dice} = 1 - \frac{2\sum_{x\in\Omega} p_l(x)g_l(x)}{\sum_{x\in\Omega} p_l^2(x) + \sum_{x\in\Omega} g_l^2(x)} \tag{1}$$

The focal loss function is used to help mitigate the imbalance between foreground and background classes during training. The focal loss is derived from the cross entropy (CE) loss and introduces a focusing parameter $\gamma$ that helps increase the importance of correcting misclassified examples [34]. $L_{focal}$ can be written as

$$L_{focal} = -\sum_{x\in\Omega}\left(\alpha\bigl(1 - p_l(x)\bigr)^{\gamma} g_l(x) \log p_l(x) + (1-\alpha)p_l^{\gamma}(x)(1 - g_l(x))\log(1 - p_l(x))\right) \tag{2}$$

Where the weighting factor $\alpha \in [0,1]$, focusing parameter $\gamma \geq 0$, $g_l(x)$ and $p_l(x)$ are label and estimated probability vectors, respectively. In our experimental designs, $\alpha = 0.25$ and $\gamma = 2$ works best in practice [34].

### 3. Results

*3.1 Patient Demographics*

Our dataset comprised of images from 50 patients (20 control and 30 diabetic retinopathy eyes). Subjects and diabetic patients with and without DR were recruited from the UIC retina clinic. The patients present in this study are representative of a university population of diabetic patients who require clinical diagnosis and management of DR. Two board-certified retina specialists classified the patients based on the severity of DR according to the Early Treatment Diabetic Retinopathy Study (ETDRS) staging system. All patients underwent complete anterior and dilated posterior segment examination (JIL, RVPC). All control OCTA images were obtained from healthy volunteers that provided informed consent for OCT/OCTA imaging. All subjects underwent OCT and OCTA imaging of both eyes (OD and OS). The images used in this study did not include eyes with other ocular diseases or any other pathological features in their retina such as epiretinal membranes and macular edema. Additional exclusion criteria included eyes with prior history of intravitreal injections, vitreoretinal surgery or significant (greater than a typical blot hemorrhage) macular hemorrhages. Validation dataset comprised of healthy volunteers that provided informed consent for OCT/OCTA imaging.

*3.2 Classification Evaluation*

The AV-Net achieved an average accuracy of 86.75% (86.71% and 86.80% respectively for artery and vein) on the test data and a mean IOU was 70.72%, and F1-score of 82.81%. The results of our study are summarized in Table 1.

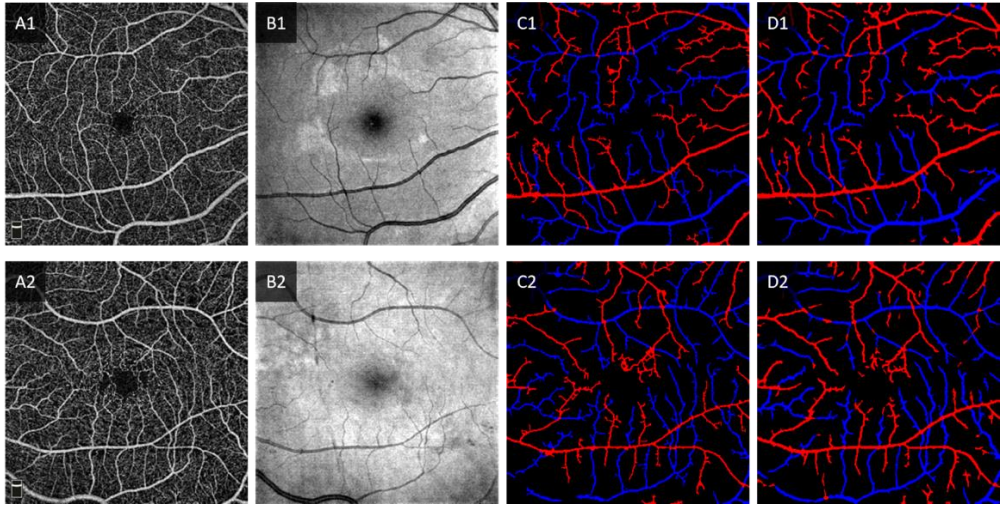

Figure 2. Examples of control and DR (top and bottom, respectively) (a) input OCTA, (b) enface OCT, (c) the ground truth, and (d) the predicted AV-maps.

In this study, the dataset was comprised of enface OCT and OCTA from healthy control and NPDR patients, example qualitative inputs and predicted AV maps from our network are shown in Fig. 2. Qualitatively, the overall vessel segmentation is robust. For the individual vessel classifications, we can see that AV-Net performs well, however, since AV-Net performs pixel-wise semantic classification, there are some regions of the vessels that have misclassification.

Table 1. AV classification performance with the AV-Net

| Cross Validation | Accuracy | F1 | IOU |
| --- | --- | --- | --- |
| Artery | 86.705 ± 1.087 | 82.761 ± 1.677 | 70.658 ± 2.404 |
| Vein | 86.798 ± 1.174 | 82.850 ± 1.666 | 70.781 ± 2.399 |
| Average | 86.751 ± 1.126 | 82.805 ± 1.670 | 70.719 ± 2.399 |

## 4. Discussion

In summary, we have demonstrated the AV-Net for fully automated AV classification in OCTA. The AV-Net achieved an average accuracy of 86.75% (86.71% and 86.80% respectively for artery and vein) on the test data and a mean IOU was 70.72%, and F1-score of 82.81%.

Differential AV analysis is known to be valuable for quantifying subtle microvascular changes and distortions due to retinopathies. Incorporating AV classification capability into the clinical imaging devices would enhance the diagnostic ability and quantitative power of OCTA. Previous studies exploring the use of deep learning for AV classification have been primarily focused on traditional fundus photography. Xu et al. adapted a UNet for AV classification using publicly available fundus datasets, such as DRIVE and INSPIRE, and achieved high accuracy [35]. Similarly, Meyer et al. employed deep learning using a patch-wise prediction strategy and included regularization techniques such as dropout and batch normalization [36]. To our knowledge, this is the first study to employ deep learning for AV classification in OCTA.

In this study, we employed an FCN, based on the UNet architectures. In a previous study, Ronneberger et.al. [27] have shown the use of long skip connections, that can help the network localize high resolution features, thereby a more precise output. In AV-Net, we employ dense blocks that utilize short skip connections. These short skip connections encourage the network to reuse features, making the model more compact. In comparison to other networks such as VGG16, AV-Net is a 5 times deeper network (having more convolutional layers) but the number of parameters is significantly smaller (approximately 17 times less). Having deeper network enables more learning capability, whereas smaller number of parameters means less computational burden. By leveraging both long and short skip connections, we are able to train our AV-Net for robust AV classification.

The input of the AV-Net consists of both enface OCT and OCTA. While OCTA does provide highly detailed vasculature maps, the arteries and veins are indistinguishable from each other by OCTA information itself. On the other hand, OCT retains reflectance information to differentiate artery and vein [25]. By combining both images, the FCN can learn the intensity information from the OCT and the highly detailed vasculature from the OCTA. Employing both OCT and OCTA is also convenient since they are from same OCT data volume and OCTA is reconstructed based on OCT processing. Therefore, using enface OCT and OCTA as 2-channel input of the AV-Net requires no pre-processing and image registration.

The results of the cross-validation study revealed an adequate IOU and F1 score. Qualitatively AV-Net has good vessel segmentation and AV classification performance. However, the predicted AV maps do appear more dilated compared to the ground truths. There are notable areas of misclassification, i.e., at vessel cross points. Future improvements to AV-Net could include developing a dataset with ground truth for vessel crossings. Additional validation with enlarged datasets from different OCTA devices will be required to pursue clinical deployments of the AV-Net for differential AV analysis.

## 5. Conclusion

The AV-Net has been demonstrated for fully automated AV classification in OCTA. The AV-Net is based on one FCN with modified U-shaped CNN architecture. A multi-modal training process was involved to include both enface OCT and OCTA for robust AV classification, and a transfer learning procedure was integrated to compensate for the limited size of OCTA dataset. By incorporating transfer learning and multi-modal training, the AV-Net achieved an accuracy of 86.75% for robust AV classification.

## Funding

This research was supported in part by NIH grants R01 EY030101, R01EY029673, R01 EY023522, R01EY030842, P30 EY001792; by unrestricted grant from Research to Prevent Blindness; by Richard and Loan Hill endowment; by Illinois society to prevent blindness grant.

## Disclosure

No competing interest exists for any author.